\documentclass{moriond}
\usepackage{amssymb}
\usepackage{sidecap}

\def\Journal#1#2#3#4{{#1} {\bf #2}, #3 (#4)}

\def\NPA{{\em Nucl. Phys.} A}

\def\PLB{{\em Phys. Lett.}  B}
\def\PRL{\em Phys. Rev. Lett.}

\def\PRC{{\em Phys. Rev.} C}

\def\JHEP{{\em JHEP}}
\def\JPG{{\em J. Phys.} G}
\def\JI{{\em J. Instrum.}}
\def\IJMP{{\em Int. J. Mod. Phys.}}
\def\EPJC{{\em Eur. Phys. J.} C}

\def\be{\begin{equation}}
\def\ee{\end{equation}}
\def\bea{\begin{eqnarray}}
\def\eea{\end{eqnarray}}

\newcommand{\Photo}{}

\begin{document}

\vspace*{4cm}

\title{Heavy Flavours in ALICE}

\author{ Y. Pachmayer for the ALICE Collaboration }

\address{Physikalisches Institut der Universit\"{a}t Heidelberg, 
Im Neuenheimer Feld 226, D-69120 Heidelberg, Germany}

\maketitle\abstracts{
The measurement of heavy-flavour production allows us to characterise the properties of the deconfined medium created in Pb--Pb collisions at LHC energies. pp collisions serve as a reference for Pb--Pb studies, and p--Pb collisions provide information on initial and/or final state effects related to cold nuclear matter. We report on open heavy-flavour as well as quarkonium production in p--Pb collisions at \mbox{$\sqrt{s_{\rm NN}}$ = 5.02 TeV.} The experimental data are compared with results from Pb--Pb collisions as well as with various theoretical predictions.}
\vspace{-0.8cm}
\section{Introduction}\label{sec:intro}
\vspace{-0.3cm}
The ALICE detector at the LHC is dedicated to studying the properties of the high-density, colour-deconfined state of strongly-interacting matter (Quark-Gluon Plasma) that is created in ultrarelativistic heavy-ion collisions at the LHC. \newline Heavy quarks (c, b) are excellent probes of the QGP. Due to their large mass, the quarks are
produced in initial hard scattering processes occurring on a short time scale ($\tau \approx$ 1/(2$m_{\rm q}$) $\lesssim$ 0.1 fm/$c$) compared with the QGP formation time and are thus sensitive to the full history of the collision. In particular, they enable us to study parton energy loss and its quark mass dependence. A sensitive observable for the study of the interaction of hard partons with the medium is the nuclear modification factor $R_{\rm AA}$. It is defined as the ratio of the $p_{\rm T}$ distribution of a given particle specie in heavy-ion collisions to that in pp collisions at the same centre-of-mass energy, scaled with the corresponding number of nucleon-nucleon collisions. In the 20~\% most-central Pb--Pb collisions at $\sqrt{s_{\rm NN}}$ = 2.76 TeV the D-meson $R_{\rm AA}$ shows a suppression of a factor of 3-4 for \mbox{$p_{\rm T} \ge$ 5 GeV/$c$~\cite{ALICEDmeson}.} \newline
Quarkonium states are expected to be suppressed in the QGP, due to colour screening of the
force which binds the $c\bar{c}$ (or $b\bar{b}$) state. The measurement of J/$\psi$ production in heavy-ion collisions was therefore proposed in 1986 by Matsui and Satz as a probe to study the onset of \mbox{de-confinement \cite{satz}.} When studying Pb--Pb collisions at LHC energies, it is expected that the abundant production of charm quarks in the initial state leads to additional charmonium generation from (re-)combination of $c$ and $\bar{c}$ quarks along the collision history and/or at \mbox{hadronisation \cite{pbm,thews},} resulting in an enhancement of the J/$\psi$ yield. Indeed the J/$\psi$ $R_{\rm AA}$ at LHC \cite{ALICEjpsi} shows less suppression than similar measurements performed in Au--Au collisions at $\sqrt{s_{\rm NN}}$ = 200 GeV/$c$ \cite{jpsiphenix}. \newline
In order to better interpret the aforementioned results in Pb--Pb collisions, it is essential to study heavy-flavour production in p--Pb collisions. This gives us access to nuclear effects, which are not related to the formation of a deconfined medium.
\vspace{-0.3cm}
\section{Heavy-flavour measurements in ALICE}\label{sec:meas}
\vspace{-0.3cm}
ALICE is well suited to studying open charm and beauty hadrons, as well as charmonium and bottomonium states due to its excellent particle identification, momentum resolution and, in the central barrel, its track impact parameter resolution \cite{ALICEexperiment}. In this paper heavy-flavour production in p--Pb collisions at $\sqrt{s_{\rm NN}}$ = 5.02 TeV is presented in the following channels:
\vspace{-0.3cm}
\begin{itemize}
\item Open charm and beauty: reconstruction of electrons from semi-electronic decays: \newline
\noindent D, B $\rm \rightarrow e + X$ in $\rm -1.06 < \textit{y}_{\rm cms} < 0.14$ 
\vspace{-0.3cm}
\item Open charm: fully reconstructed hadronic decays using displaced vertex identification: 
$\rm D^{0} \rightarrow K^{-} \pi^{+}$, $\rm D^{+} \rightarrow K^{-} \pi^{+}\pi^{+}$, $\rm D^{*+} \rightarrow D^{0} \pi^{+}$, $\rm D^{+}_{s} \rightarrow K^{+} K^{-} \pi^{+}$, and their respective charge conjugates in $\rm -0.96 < \textit{y}_{\rm cms} < 0.04$ 
\vspace{-0.3cm}
\item J/$\psi$ $\rm \rightarrow e^+ + e^-$ down to $p_{\rm T}$ = 0 in  $\rm -1.37 <\textit{y}_{\rm cms} < 0.43$ 
\vspace{-0.3cm}
\item J/$\psi$ $\rm \rightarrow \mu^+ + \mu^-$ down to $p_{\rm T}$ = 0 in $\rm -4.46 <\textit{y}_{\rm cms} < -2.96$ and $\rm 2.03 <\textit{y}_{\rm cms} < 3.53$ \newline
The reversal of the beam configuration (p--Pb/Pb--p) allows us to study quarkonium production with the muon spectrometer both at forward $\textit{y}_{\rm cms}$, i.e. in the p-going side, and at backward $\textit{y}_{\rm cms}$, i.e. in the Pb-going side.
\end{itemize}

\begin{figure}
\begin{minipage}{0.4\linewidth}
\centerline{\includegraphics[width=1\linewidth]{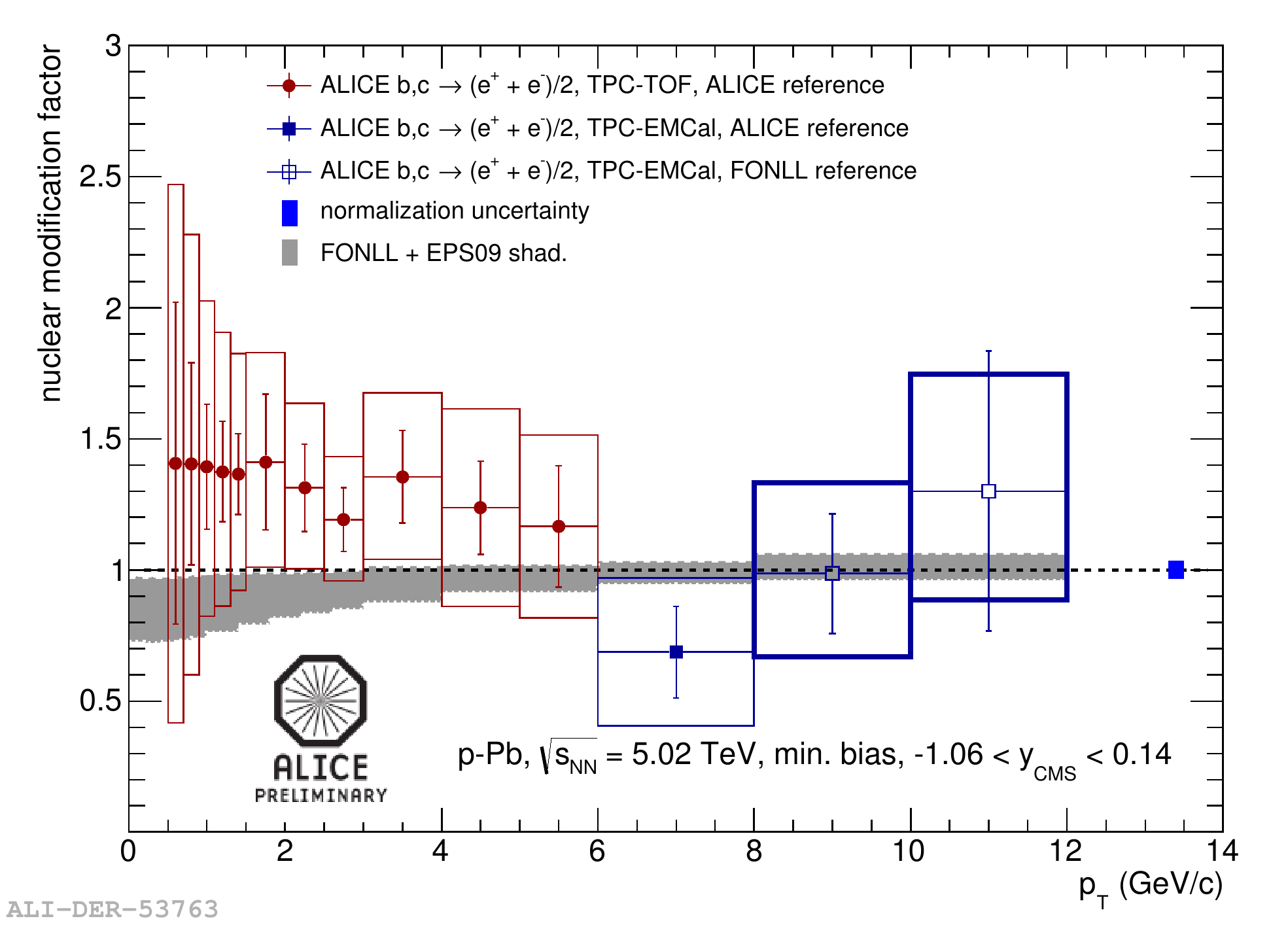}}
\end{minipage}
\hfill
\begin{minipage}{0.4\linewidth}
\centerline{\includegraphics[width=0.75\linewidth]{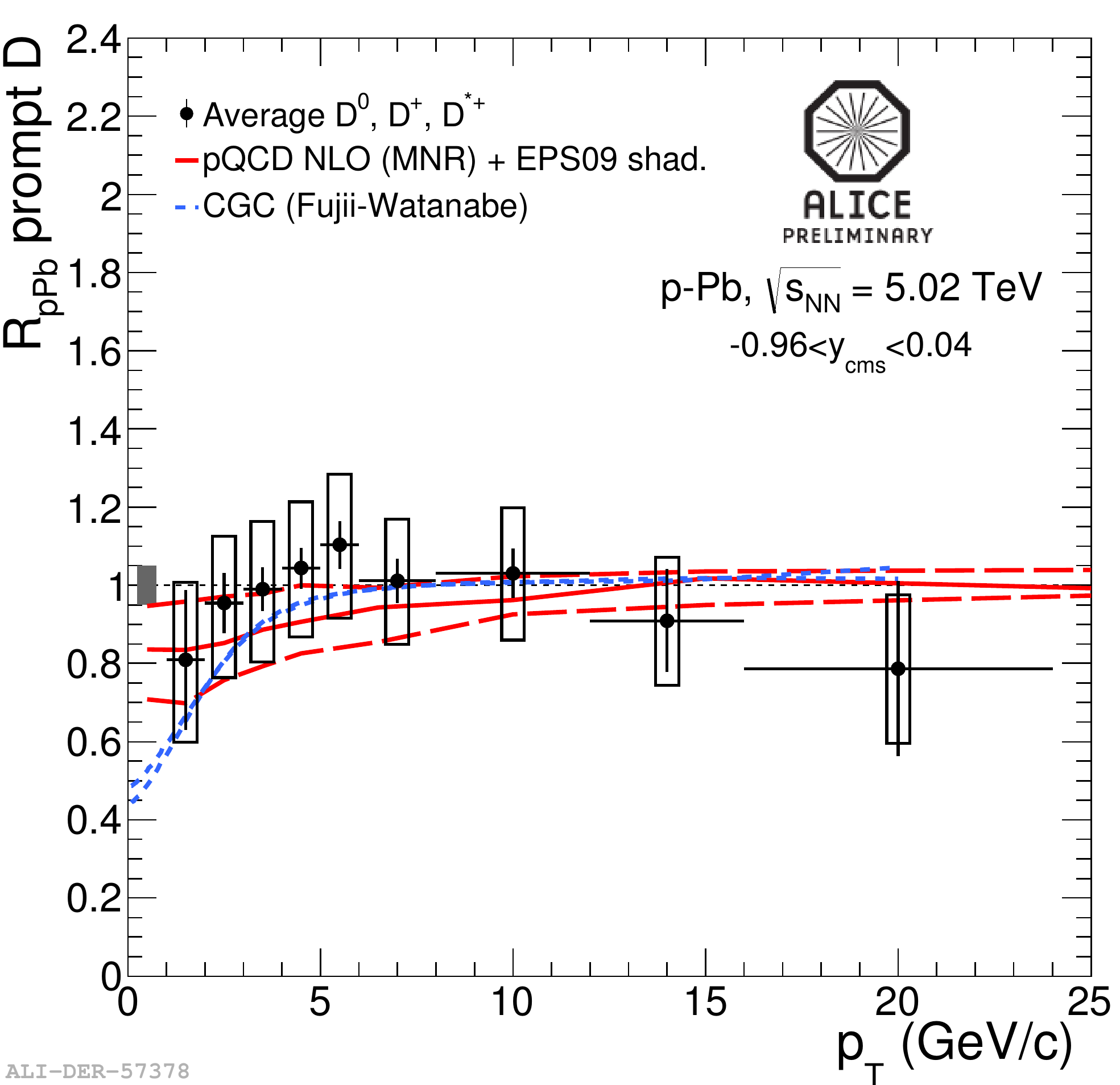}}
\end{minipage}
\hfill
\caption[]{Left: $R_{\rm pPb}$ for electrons from heavy-flavour hadron decays. Right: Average $R_{\rm pPb}$ of $D^0$, $D^+$, and $D^{*+}$.}
\label{fig:OpenHFE}
\end{figure}
\vspace{-0.5cm}
\section{Results}\label{subsec:results}
\begin{figure}
\begin{minipage}{0.33\linewidth}
\centerline{\includegraphics[width=0.95\linewidth]{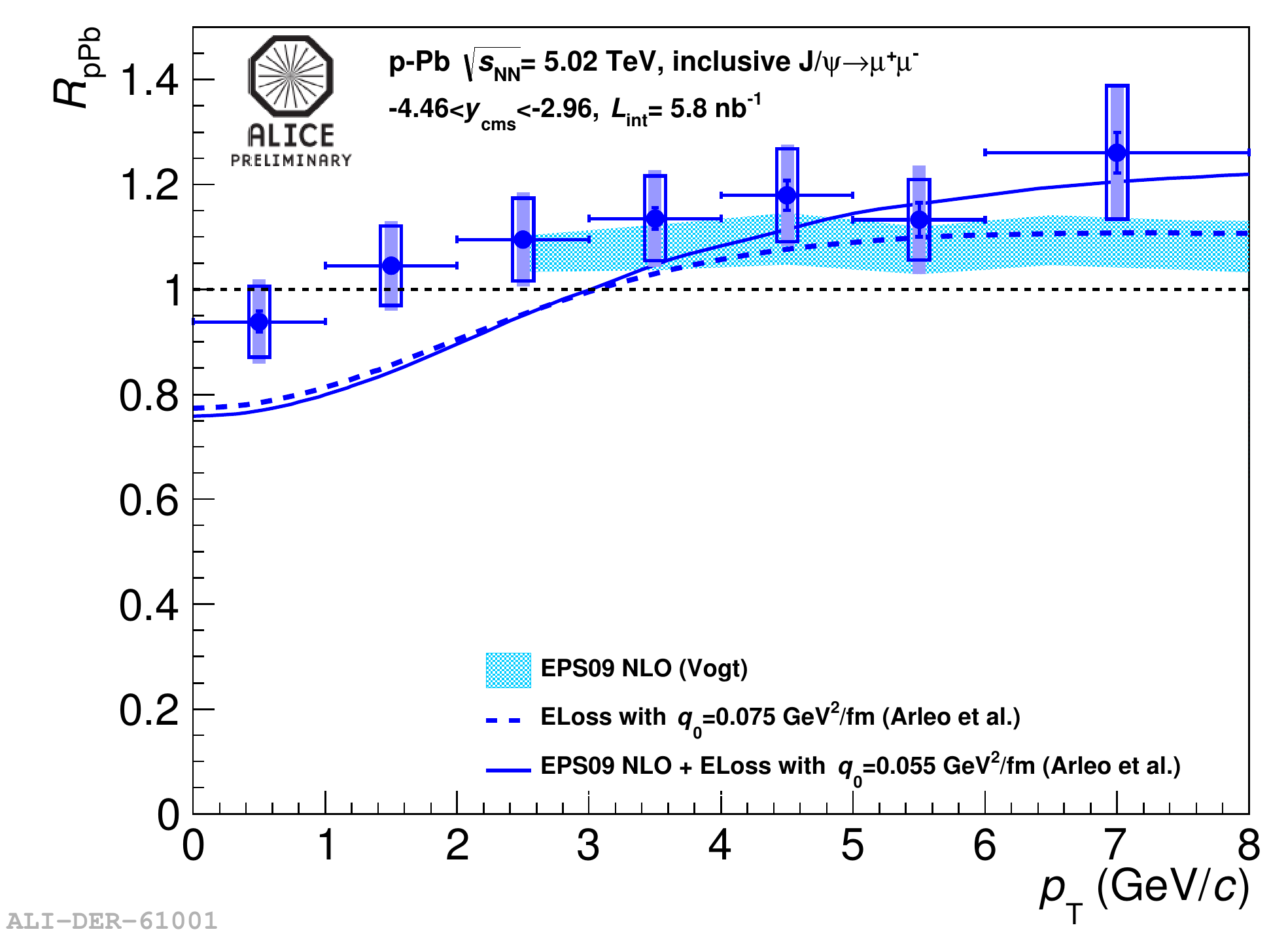}}
\end{minipage}
\hfill
\begin{minipage}{0.32\linewidth}
\centerline{\includegraphics[width=1.1\linewidth]{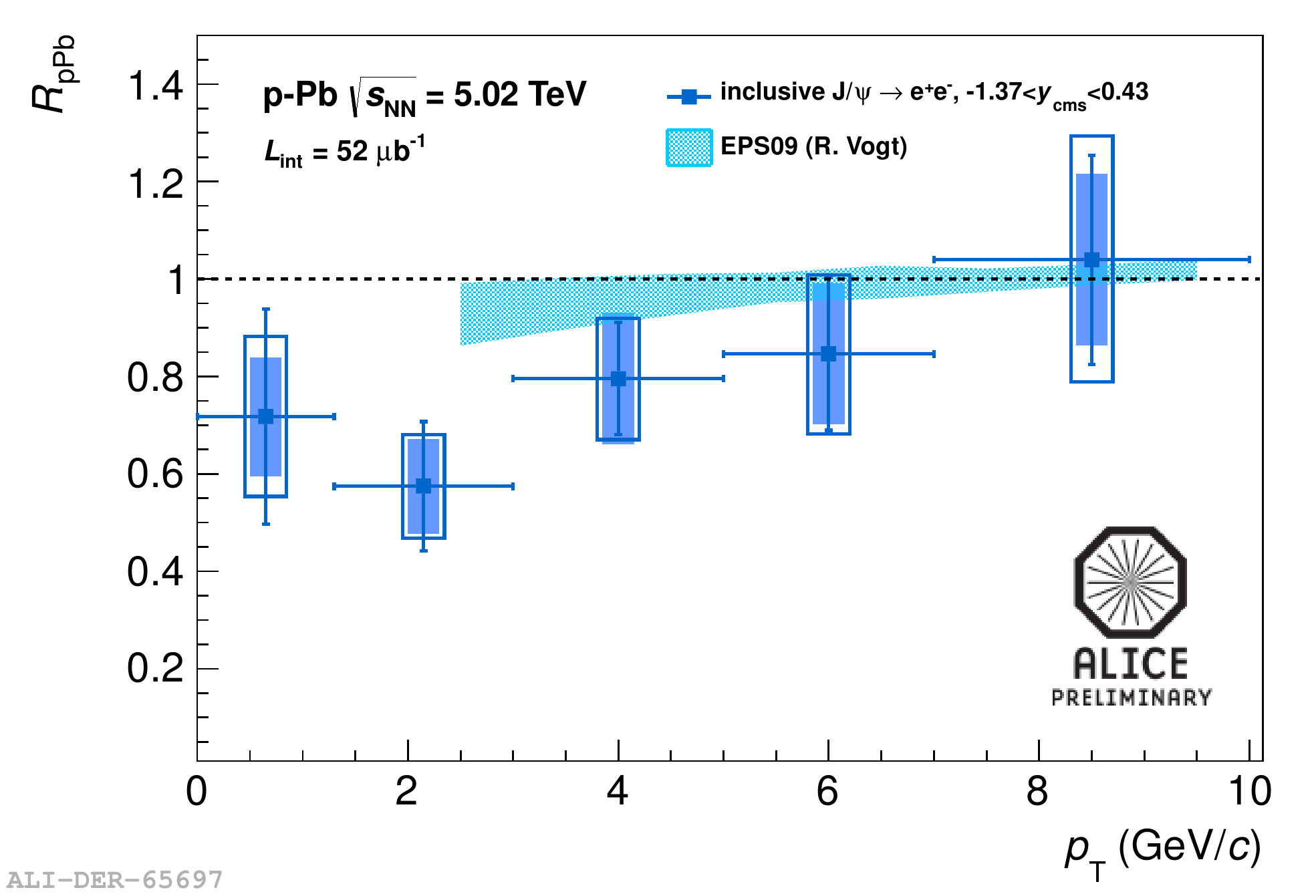}}
\end{minipage}
\hfill
\begin{minipage}{0.32\linewidth}
\centerline{\includegraphics[width=0.95\linewidth]{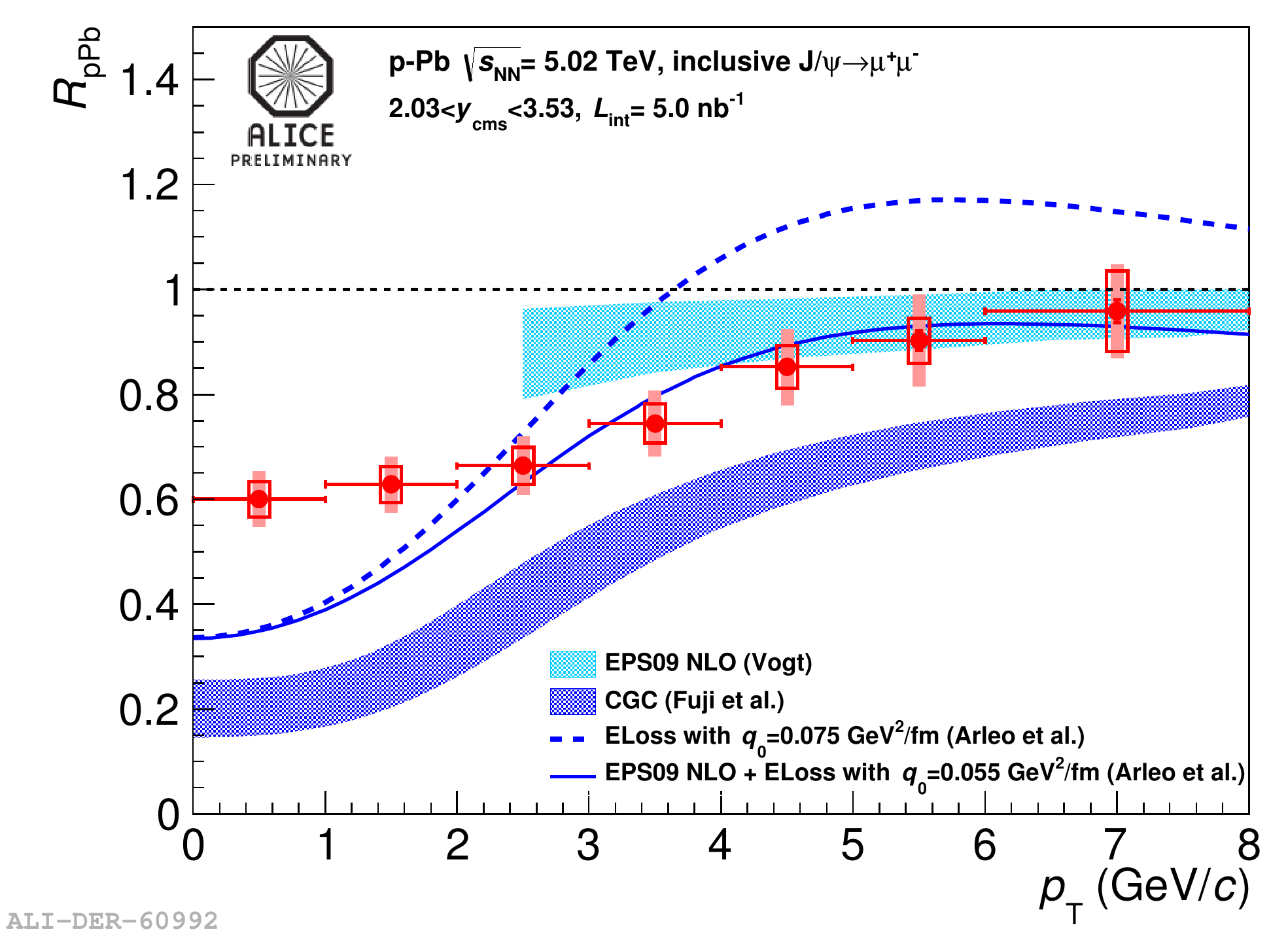}}
\end{minipage}
\caption[]{ J/$\psi$ $R_{\rm pPb}$ as a function of $p_{\rm T}$ for backward (left), mid- (center), and forward rapidity (right). The error bars, open boxes and the shaded areas
represent, respectively, the statistical, the uncorrelated and the partially correlated uncertainties. Results for various models are also shown \cite{CGC,vogt,arleo}.}
\label{fig:jpsipA}
\end{figure}

\begin{figure}
\begin{minipage}{0.45\linewidth}
\centerline{\includegraphics[width=0.83\linewidth]{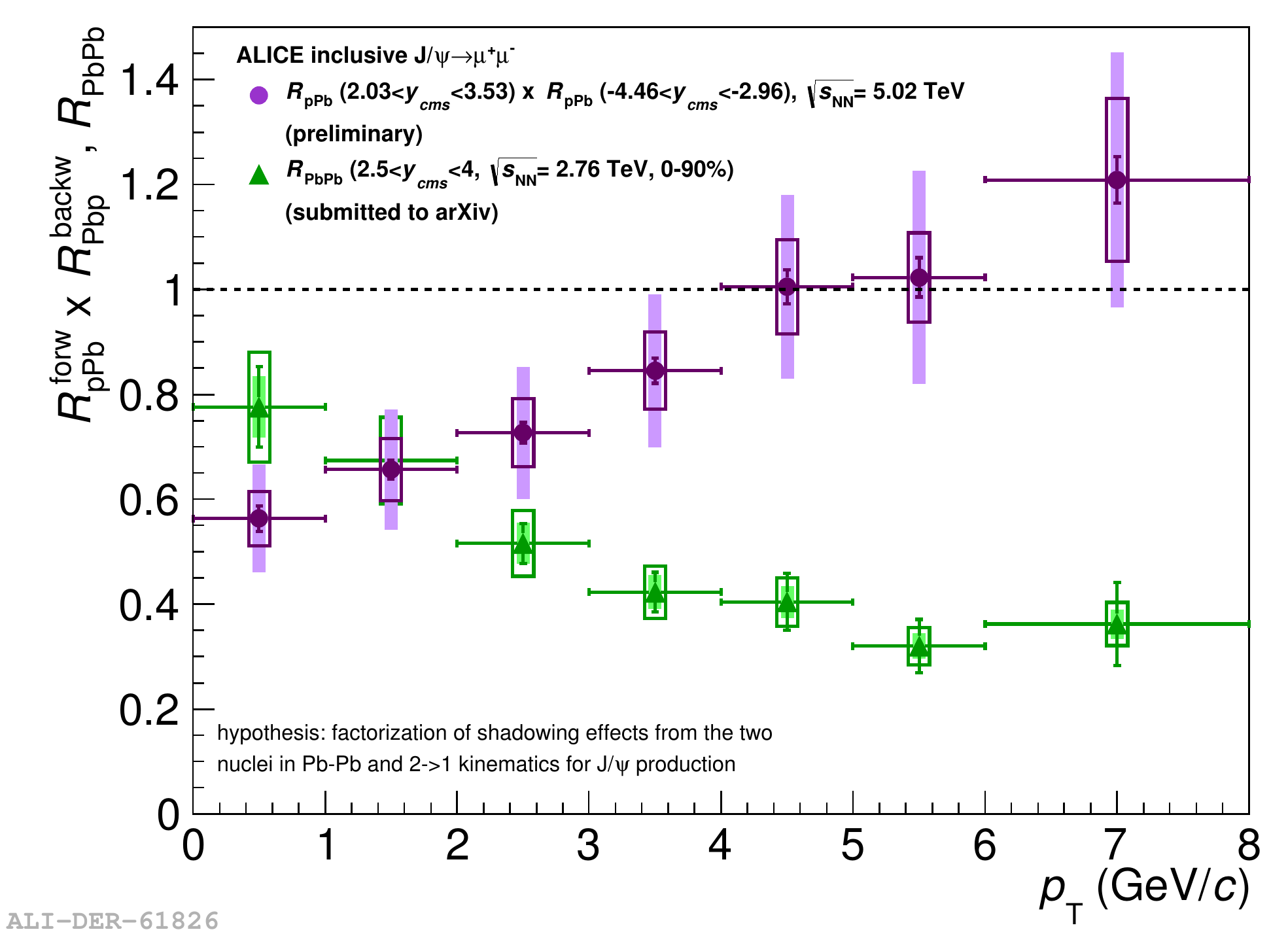}}
\end{minipage}
\hfill
\begin{minipage}{0.45\linewidth}
\centerline{\includegraphics[width=0.9\linewidth]{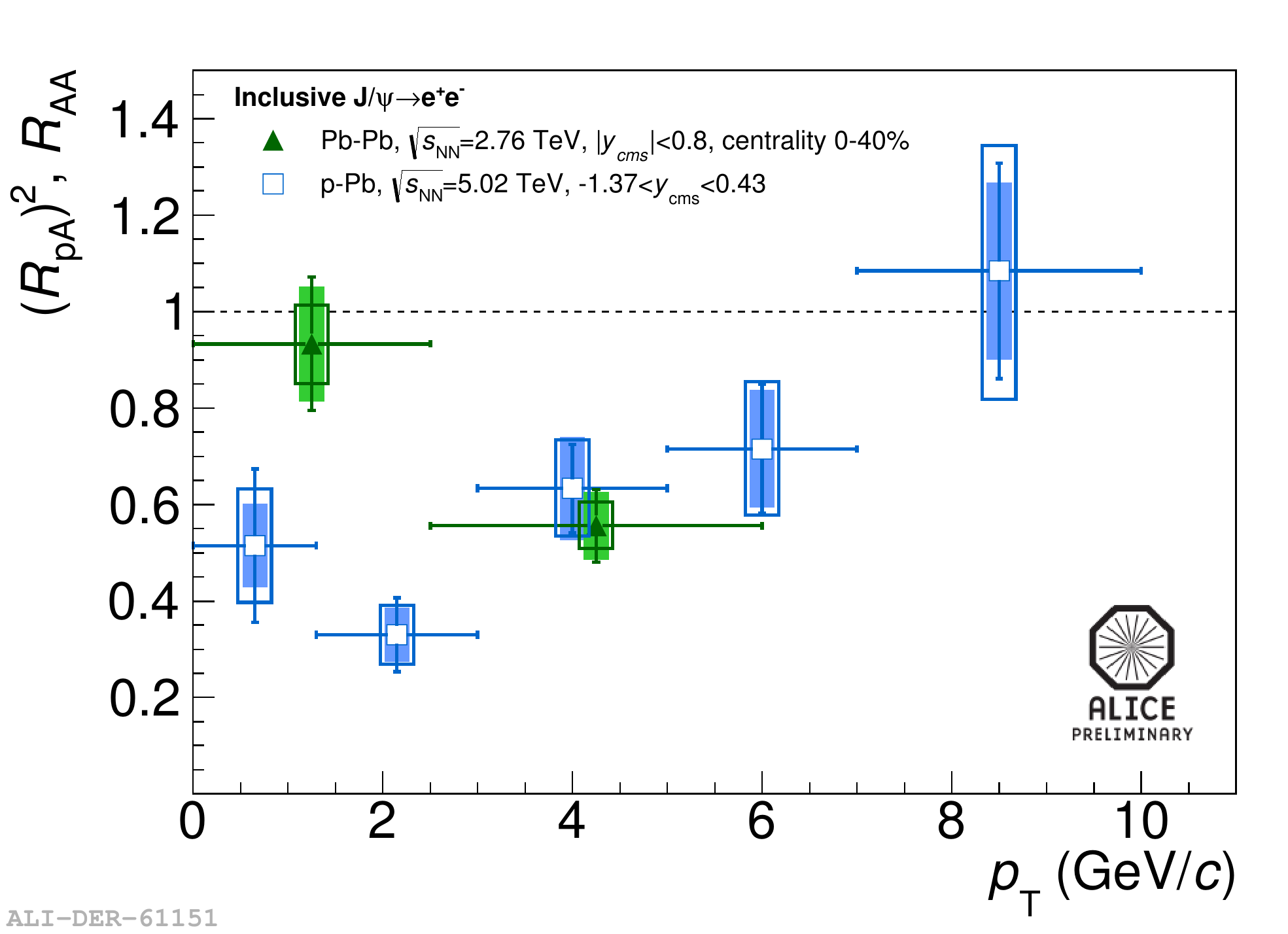}}
\end{minipage}
\hfill
\caption[]{Comparison of J/$\psi$ $R_{\rm pPb}^{\rm forward} \times R_{\rm pPb}^{\rm backward}$ with $R_{\rm AA}$ at forward rapidity (left) and $\left( R_{\rm pPb}^{\rm mid-rapidity}\right)^{2}$ with $R_{\rm AA}$ at mid-rapidity (right). The centrality of the Pb--Pb collisions is indicated in the figures.}
\label{fig:jpsipAAA}
\end{figure}

\vspace{-0.3cm}
The nuclear modification factor of electrons from heavy-flavour hadron decays in p--Pb collisions at $\sqrt{s_{\rm NN}}$ = 5.02 TeV is shown in Fig.~\ref{fig:OpenHFE} (left). The pp reference spectrum is obtained by scaling the measured $p_{\rm T}$-differential cross section of electrons from heavy-flavour hadron decays in pp collisions at $\sqrt{s}$ = 7 TeV to \mbox{5.02 TeV} based on FONLL calculations. 
The $R_{\rm pPb}$ is consistent with unity within uncertainties. Predictions based on perturbative QCD calculations including initial state effects (FONLL \cite{FONLLcalc} + EPS09 \cite{MNR} nuclear PDFs) agree with the data within uncertainties. \newline
The $R_{\rm pPb}$ of the $D^0$, $D^+$, $D^{*+}$ and $D^+_s$ mesons are compatible with one another and consistent with unity; the average value calculated from the three non-strange D-meson species is shown in Fig.~\ref{fig:OpenHFE} (right). Next-to-leading-order pQCD calculations using the MNR framework with EPS09 nuclear PDFs parameterisation \cite{MNR} as well as Colour-Glass Condensate (CGC) \mbox{predictions \cite{CGC}} describe well the data. The result from p--Pb collisions indicates that the large suppression observed in Pb--Pb collisions (see Sect \ref{sec:intro}) is a final state effect, due to in-medium charm quark energy loss. \newline
The $R_{\rm pPb}$ of inclusive J/$\psi$ mesons integrated over $p_{\rm T}$ shows a suppression by a factor of $\sim$1.4 at forward rapidity, while the measurement at backward rapidity is consistent with unity \cite{ALICEjpsipPb}. The effect of non-prompt J/$\psi$ production on the $R_{\rm pPb}$ was estimated to be at most 14~\% (25~\%) at low (high) $p_{\rm T}$. The dependence of $R_{\rm pPb}$ on transverse momentum is shown in Fig.~\ref{fig:jpsipA} for three rapidity intervals. At backward rapidity the $R_{\rm pPb}$ shows a small $p_{\rm T}$ dependence and is close to unity. Even considering the large uncertainties, the $R_{\rm pPb}$ at mid-rapidity tends to increase with $p_{\rm T}$. At forward rapidity, the $R_{\rm pPb}$ clearly rises with $p_{\rm T}$ and is consistent with unity for \mbox{$p_{\rm T} \ge$ 5 GeV/$c$.} Cold nuclear matter effects (CNM) are small in all rapidity domains for \mbox{$p_{\rm T} \ge$ 4 GeV/$c$.} \newline Various model calculations for prompt J/$\psi$ production were compared with the experimental data. The first calculation by Vogt was obtained in the Colour Evaporation Model (CEM) at next-to-leading order employing EPS09 shadowing parameterisations \cite{vogt}. It reproduces the $p_{\rm T}$ dependence in all three rapidity domains in the $p_{\rm T}$-range provided. The coherent parton energy loss model \cite{arleo} with and without EPS09 shadowing as an additional nuclear effect describes the measurement for $p_{\rm T} \geq$ 2 GeV/$c$. A calculation in the CGC framework \cite{CGC}, combined with a CEM production model, overestimates the suppression at forward rapidity and is thus disfavoured by the data. \newline
\noindent Assuming a 2 $\rightarrow$ 1 production mechanism for the J/$\psi$ meson ($gg \rightarrow$ J/$\psi$), the Bjorken-\textit{x} values of the lead nucleus match in p--Pb and Pb--Pb collisions. Further, assuming shadowing as the main nuclear effect and factorisation of CNM effects, an expectation for $R_{\rm AA}$ due to CNM effects only can be calculated by considering $R_{\rm pPb}^{\rm forward} \times R_{\rm pPb}^{\rm backward}$ ($\left(R_{\rm pPb}^{\rm mid-rapdidity}\right)^{2}$). The resulting quantity is shown as a function of $p_{\rm T}$ in comparison with the $R_{\rm AA}$ measured at forward (mid-) rapidity in Fig.~\ref{fig:jpsipAAA}. The effects from the extrapolated shadowing are small at \mbox{$p_{\rm T} \ge$ 7 (4) GeV/$c$} at mid- (forward) rapidity. At low  $p_{\rm T}$ the result of $R_{\rm AA}$ is equal to or enhanced compared with the CNM expectation from p--Pb measurements. This behaviour is in qualitative agreement with theoretical calculations taking into account non-primordial J/$\psi$ production \cite{pbm,thews}. 

\begin{figure}
\begin{minipage}{0.45\linewidth}
\centerline{\includegraphics[width=0.85\linewidth]{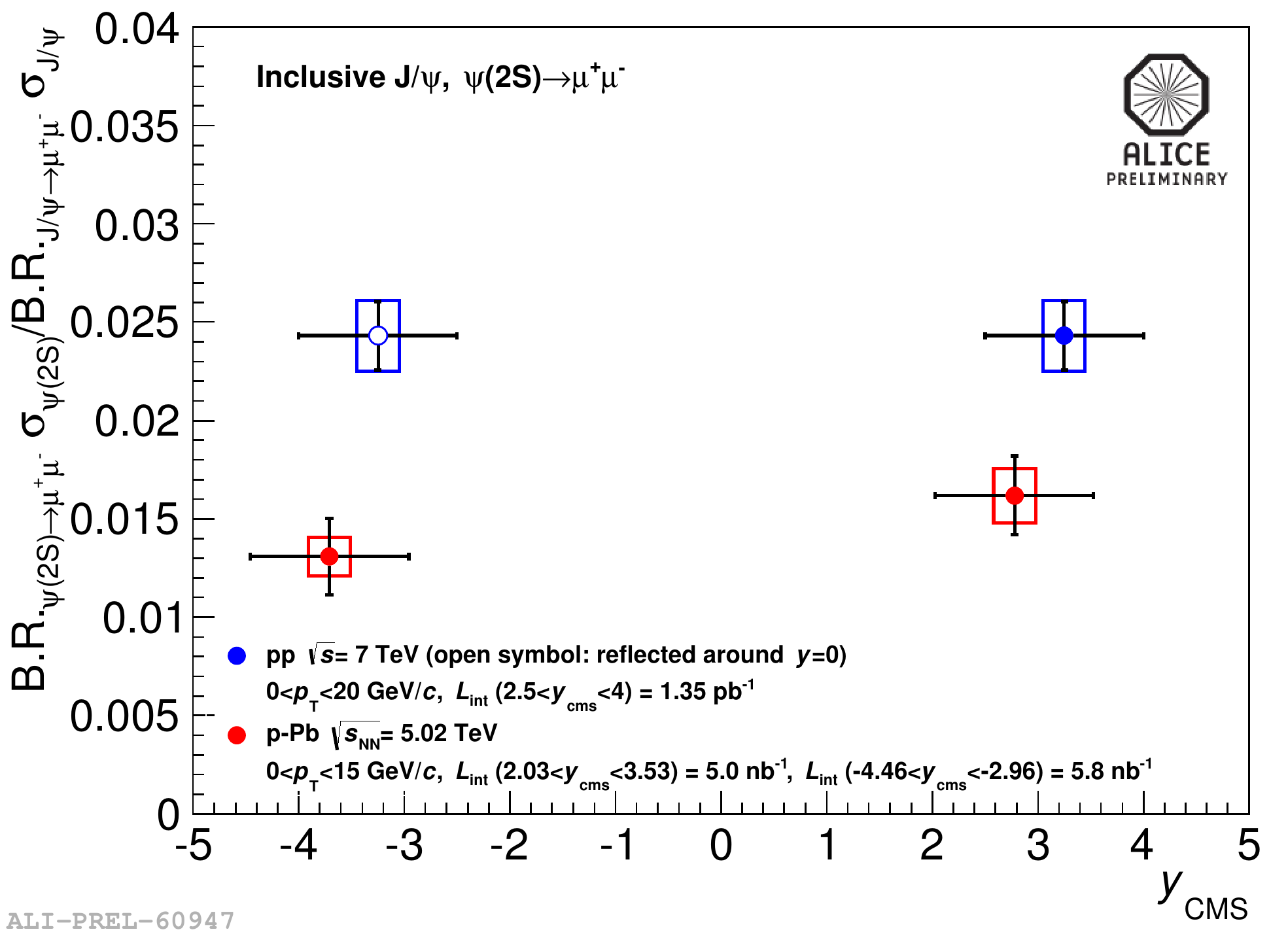}}
\end{minipage}
\hfill
\begin{minipage}{0.44\linewidth}
\centerline{\includegraphics[width=0.89\linewidth]{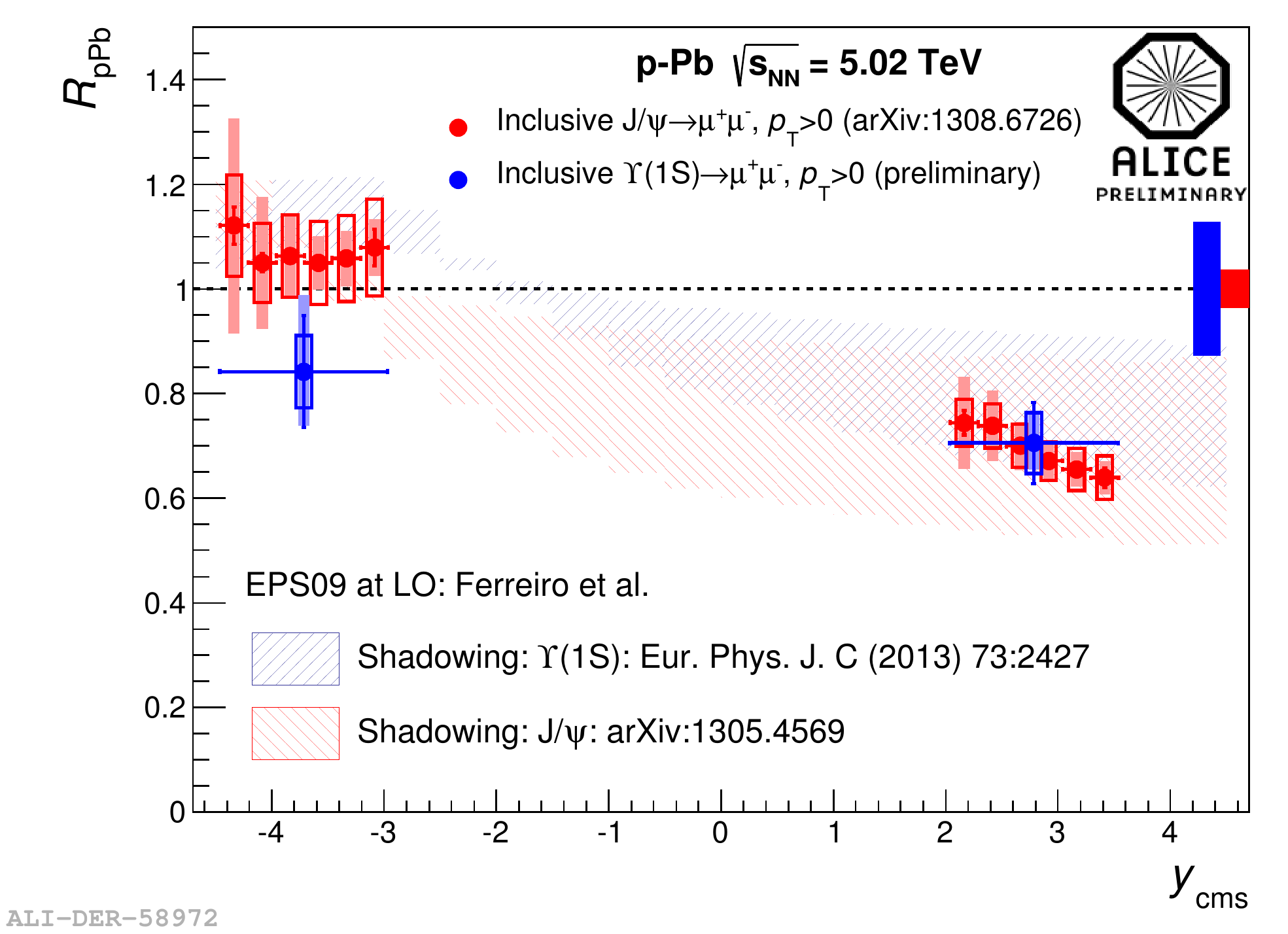}}
\end{minipage}
\hfill
\caption[]{Left: $\psi$(2S)/J/$\psi$ vs rapidity for pp collisions at $\sqrt{s}$ = 7 TeV  and p--Pb collisions at $\sqrt{s_{\rm NN}}$ = 5.02 TeV. Right: Inclusive $\Upsilon$(1S) $R_{\rm pPb}$ as a function of rapidity compared with the J/$\psi$ $R_{\rm pPb}$.}
\label{fig:psi2upsilon}
\end{figure}

\noindent Analogous to the J/$\psi$ measurement, a $\psi$(2S) $\rightarrow \mu^+ \mu^-$ analysis was conducted. Figure \ref{fig:psi2upsilon} (left) shows the ratio of $\psi$(2S) to J/$\psi$ as a function of rapidity for p--Pb collisions. The depicted pp $\psi$(2S)/J/$\psi$ ratio was obtained at a different energy ($\sqrt{s}$ = 7 TeV) and in a slightly different rapidity region (2.5  $\le y_{\rm cms} \le$ 4). The ratio $\psi$(2S)/J/$\psi$ shows a strong decrease in p--Pb compared with pp collisions. This is not expected in models of initial state effects, such as shadowing, coherent energy loss or the CGC-based model (see above), which all predict the same cold nuclear matter effects on J/$\psi$ and $\psi$(2S) and, as a consequence, the same $\psi$(2S)/J/$\psi$ ratio in pp and p--Pb collisions. The larger suppression observed for $\psi$(2S) compared with J/$\psi$ could therefore be due to a final state effect. \newline
\noindent $\Upsilon$(1S) $\rightarrow \mu^+ \mu^-$ was also measured at backward and forward rapidity. The respective $R_{\rm pPb}$ is shown in Fig.~\ref{fig:psi2upsilon} (right) as a function of $p_{\rm T}$. The uncertainties are dominated by the energy interpolation of the pp cross section and the limited statistics. At forward rapidity the $\Upsilon$(1S) and J/$\psi$ are suppressed to a similar extent. The model calculation by Ferreiro \cite{ferre} (EPS09 nPDF associated with a Colour Singlet Model at LO) is in fair agreement with data within uncertainties. 

\vspace{-0.2cm}
\section{Conclusions}\label{subsec:concl}
\vspace{-0.3cm}
Heavy-flavour production was presented for p--Pb collisions at $\sqrt{s_{\rm NN}}$ = 5.02 TeV and compared with the respective measurements in pp and Pb--Pb collisions. \newline The open heavy-flavour results are well described by pQCD calculations including shadowing predictions and confirm that the suppression seen in central Pb--Pb collisions is a final state effect, due to in-medium parton energy loss. \newline
\noindent The J/$\psi$ $R_{\rm pPb}$ was measured as a function of rapidity and transverse momentum. At forward rapidity the measurement supports strong shadowing and/or the coherent energy loss model. Using the measurements in p--Pb, an expectation of the contribution of CNM effects in \mbox{Pb--Pb} was calculated. At low $p_{\rm T}$, the observed $R_{\rm AA}$ is higher than the calculated expectation, thus reinforcing the interpretation of an additional (re-)combination mechanism for J/$\psi$ production in Pb--Pb.
$\psi$(2S) is suppressed by up to 45\% relative to J/$\psi$ at backward rapidity, which is probably caused by a final state effect. The $\Upsilon$(1S) $R_{\rm pPb}$ shows a similar suppression as for J/$\psi$.

\end{document}